\documentclass[a4paper,11pt]{article}

\usepackage[parfill]{parskip}  
\usepackage[toc,page]{appendix}
\usepackage[english]{babel}
\usepackage{amsmath}
\usepackage{amsfonts}
\usepackage{amssymb}
\usepackage{amsthm}
\usepackage[pdftex]{graphicx}
\usepackage{epstopdf}
\usepackage{enumitem}
\usepackage{geometry}
\usepackage{hyperref}
\usepackage{tikz}
\usepackage[all]{xy}
\usepackage{multirow}
\usepackage{anysize}
\usepackage{color}
\usepackage{pdflscape}
\usepackage{caption}
\usepackage{lineno}

\newcommand{\beq}{\begin{equation}}
\newcommand{\eeq}{\end{equation}}
\newcommand{\commentout}[1]{}

\def\wl{\par \vspace{\baselineskip}}

\marginsize{3.2cm}{3.2cm}{3.2cm}{3.2cm}

\title{Social setting, intuition, and experience in lab experiments interact to shape cooperative decision-making}

\author{Valerio Capraro$^1$ \& Giorgia Cococcioni$^2$}

\begin{document}

\maketitle

\begin{center}
$^1$Center for Mathematics and Computer Science (CWI), 1098 XG, Amsterdam,  The Netherlands. $^2$Department of Political Sciences, LUISS Guido Carli, 00197, Roma, Italy. Contact author: V.Capraro@cwi.nl\wl
\emph{Forthcoming in Proceedings of the Royal Society B: Biological Sciences}
\end{center}

\pagebreak

\begin{abstract}
Recent studies suggest that cooperative decision-making in one-shot interactions is a history-dependent dynamic process: promoting intuition versus deliberation has typically a positive effect on cooperation (dynamism) among people living in a cooperative setting and with no previous experience in economic games on cooperation (history-dependence). Here we report on a lab experiment exploring how these findings transfer to a non-cooperative setting. We find two major results: (i) promoting intuition versus deliberation has no effect on cooperative behavior among inexperienced subjects living in a non-cooperative setting; (ii) experienced subjects cooperate more than inexperienced subjects, but only under time pressure. These results suggest that cooperation is a learning process, rather than an instinctive impulse or a self-controlled choice, and that experience operates primarily via the channel of intuition. In doing so, our findings shed further light on the cognitive basis of human cooperative decision-making and provide further support for the recently proposed Social Heuristics Hypothesis.
\end{abstract}


\pagebreak
\section*{Introduction}

One of the factors of the enormous success of our societies is our ability to cooperate \cite{trivers1971evolution,axelrod1981evolution,ostrom2000collective,milinski2002reputation,fehr2003nature,boyd2003evolution,perc2010coevolutionary,apicella2012social,capraro2013model,crockett2013models,rand2013human,zaki2013intuitive,capraro2014translucent,hauser2014cooperating,bone2014effect}. While in most animal species cooperation is observed only among kin or in very small groups, where future interactions are likely, cooperation among people goes far beyond the five rules of cooperation \cite{nowak2006five}: recent experiments have shown that people cooperate also in one-shot anonymous interactions \cite{wong2005dynamic,khadjavi2013prisoners,capraro2014heuristics,capraro2014benevolent,capraro2014good} and even in large groups \cite{barcelo2015group}. This poses an evolutionary puzzle: why are people willing to pay costs to help strangers when no future rewards seem to be at stake?

A growing body of experimental research suggests that cooperative decision-making in one-shot interactions is most likely a history-dependent dynamic process. \emph{Dynamic} because time pressure \cite{rand2012spontaneous,rand2014socialA,cone2014time,rand2014reflection,rand2014socialB}, cognitive load  \cite{schulz2012affect,cornelissen2011social,roch2000cognitive}, conceptual priming of intuition \cite{rand2012spontaneous,lotz2014spontaneous}, and disruption of the right lateral prefrontal cortex \cite{ruff2013changing} have all been shown to promote cooperation, providing direct evidence that automatic actions are, on average, more cooperative than deliberate actions. \emph{History-dependent} because it has been found that previous experience with economic games on cooperation and intuition interact such that experienced subjects are less cooperative than inexperienced subjects, but only under time pressure \cite{rand2014socialA} and that intuition promotes cooperative behavior only among inexperienced subjects with above median trust in the setting where they live \cite{rand2014reflection}. While this latter paper also shows that promoting intuition versus reflection has no effect among experienced subjects, its results are inconclusive with regard to people with little trust in their environment, due to the limited number of observations. More generally, the limitation of previous studies is that they have all been conducted in developed countries and so they do not allow to draw any conclusions about what happens among people from a societal background in which they are exposed to frequent non-cooperative acts.

Two fundamental questions remain then unsolved. What is the effect of promoting intuition versus deliberation among people living in a non-cooperative setting? How does this interact with previous experience with economic games on cooperative decision-making?

The first question is particularly intriguing since, based on existing theories, several alternatives are possible. The Social Heuristics Hypothesis (SHH), introduced by Rand and colleagues \cite{rand2012spontaneous,rand2014socialA} to explain the intuitive predisposition towards cooperation described above ``posits that cooperative decision making is guided by heuristic strategies that have generally been successful in one's previous social interactions and have, over time, become internalized and automatically applied to social interactions that resemble situations one has encountered in the past. When one encounters a new or atypical social situation that is unlike previous experience, one generally tends to rely on these heuristics as an intuitive default response. However, through additional deliberation about the details of the situation, one can override this heuristic response and arrive at a response that is more tailored to the current interaction'' \cite{cone2014time}. Then, according to the SHH, inexperienced subjects living in a non-cooperative setting should bring their non-cooperative strategy (learned in the setting where they live) in the lab as a default strategy. These subjects are then predicted to act non-cooperatively both under time pressure, because they use their non-cooperative default strategy, and under time delay, because defection is optimal in one-shot interactions. 

However, this is not the only possibility. Several studies have shown that patients who suffered ventromedial prefrontal cortex damage, which causes the loss of emotional responsiveness, are more likely to display anti-social behavior  \cite{damasio1990individuals,bechara1994insensitivity,bechara1996failure}.  These findings support the interpretation that intuitive emotions play an important role in pro-social behavior and form the basis of Haidt's Social Intuition Model (SIM) according to which moral judgment is caused by quick moral intuitions and is followed (when needed) by slow, ex post facto, moral reasoning  \cite{haidt2001emotional}. While the SIM does not make any prediction on what happens in the specific domain of cooperation, it would certainly be consistent with a general intuitive predisposition towards cooperation, mediated by positive emotions, and independent of the social setting in which an individual is embedded.

A third alternative is yet possible. Motivated by work suggesting that people whose self-control resources have been taxed tend to cheat more \cite{mead2009too,gino2011unable} and be less altruistic  \cite{halali2013all,xu2012too,achtziger2014money}, it has been argued that self-control plays an important role in overriding selfish impulses and bringing behavior in line with moral standards. This is consistent with Kohlberg's rationalist approach  \cite{kohlberg1963development}, which assumes that moral choices are guided by reason and cognition: as their cognitive capabilities increase, people learn how to take the other's perspective, which is fundamental for pro-social behavior. This rationalist approach makes the explicit prediction that promoting intuition always undermine cooperation. 

\commentout{
While this is inconsistent with the aforementioned results, it is not difficult to imagine a scenario in which the rationalist approach interacts with the SHH: cooperation may have emerged after deliberation from a neutral or non-cooperative setting, giving rise to rational cooperative societies, whose members have internalized cooperation as a default strategy and use it when they encounter a new or atypical setting. Some of them, after deliberation, may switch to defection because they do not perceive a moral obligation in behaving cooperatively in one-shot anonymous interactions. Seen in this light, the SHH is not inconsistent with Kohlberg's rationalist approach. However, this tentative to reconcile the rationalist approach with the experimental data makes clear predictions on what should happen in a non-cooperative setting: whatever the level of experience of a participant is, promoting intuition should always promote non-cooperative behavior.} 

In sum, the question of how promoting intuition versus reflection affects cooperative behavior among people living in a non-cooperative setting is far from being trivial and, based on existing theories, all three possibilities (positive effect, negative effect, no effect) are, a priori, possible. 

Concerning previous experience on economic games, while the SIM and the rationalist approach do not make any prediction about its role on cooperative decision-making among people living in a non-cooperative setting, the SHH predicts that it has either a null or a positive effect driven by intuitive responses. This because experienced participants, despite their living in a non-cooperative setting, \emph{might} have internalized a cooperative strategy to be used only in experiments. Of course, the SHH does not predict that a substantial proportion of subjects have \emph{in fact} developed such a context-dependent cooperative intuition - and this is why the predicted effect is \emph{either} positive \emph{or} null. In the former case, however, the SHH predicts that the positive effect should be driven by intuitive responses, since the SHH assumes that experience operates primarily through the channel of intuition.

Here we report on an experiment aimed at clarifying these points. We provide evidence of two major results: (i) promoting intuition versus reflection has no effect on cooperation among subjects living in a non-cooperative setting and with no previous experience with economic games on cooperation; (ii) experienced subjects are more cooperative than inexperienced subjects, but only when acting under time pressure.

Taken together, these results suggest that cooperation is a learning process, rather than an instinctive impulse or a self-controlled choice, and that experience operates primarily via the channel of intuition. In doing so, they shed further light on human cooperative decision-making and provide further support for the Social Heuristics Hypothesis.

\section*{Methods}

We have conducted an experiment using the online labor market Amazon Mechanical Turk (AMT )\cite{paolacci2010running,horton2011online,rand2012promise} recruiting participants only from India. India is a particularly suited country to hire people from for our purpose: if, as many studies have confirmed  \cite{bardhan2000irrigation,dayton2000determinants,fujiie2005conditions,henrich2005economic,henrich2006costly,herrmann2008antisocial,buchan2009globalization,gachter2009reciprocity,gachter2010culture,buchan2011global,andersson2011inequalities,bigoni2014amoral}, good institutions are crucial for the evolution of cooperation, and if, as many scholars have argued \cite{das2001public,guha2007india,quah2008curbing,miklian2013corruption}, corruption and cronyism are endemic in Indian society, then residents in India are likely to have very little trust on strangers and so they are likely to have internalized non-cooperative strategies in their every-day life. One study confirms this hypothesis, by showing that spiteful preferences are widespread in the village of Uttar Pradesh and this ultimately implies residents' inability to cooperate \cite{fehr2008spite}. At the same time, according to demographic studies on AMT population \cite{ross2010who}, India is the second most active country on AMT after the US, which facilitates the procedure of collecting data. 

Participants were randomly assigned to either of two conditions: in the \emph{time pressure} condition we measured intuitive cooperation; in the \emph{time delay} condition we measured deliberate cooperation. As a measure of cooperation, we adopted a standard two-person Prisoner's dilemma (PD) with a continuous set of strategies. Specifically, participants were given an endowment of $\$0.20$, and asked to decide how much, if any, to transfer to the other participant. The amount transferred would be multiplied by 2 and earned by the other participant; the remainder would be earned by themselves, but without being multiplied by any factor. Each participant was informed that the other participant was facing the same decision problem. Participants in the time pressure condition were asked to make a decision within 10 seconds and those in the time delay condition were asked to wait for at least 30 seconds before making their choice. After making their decision, participants had to answer four comprehension questions, after which they entered the demographic questionnaire, where, along with the usual questions, we also asked ``To what extent have you previously participated in other studies like to this one (e.g., exchanging money with strangers)?'' using a 5 point Likert-scale from ``Never'' to ``Several times''. As in previous studies \cite{rand2012spontaneous,rand2014socialA,rand2014reflection}, we used the answer to this question as a measure of participant's previous experience with economic games on cooperative decision-making. As in these studies, we say that a subject is \emph{inexperienced} if he or she answered ``never'' to the above question. In the Supplementary Online Material we also report the results of a pilot, in which we measured participants' level of experience by asking them to report the extent to which they had participated in \emph{exactly} the same task before. Although the use of the word ``exactly'' may lead to confusion, with some minor differences in the details, our main results are robust to the use of this measure (see Supplementary Online Material for more details).

After collecting the results, bonuses were computed and paid on top of the participation fee ($\$0.50$). No deception was used. 

\section*{Results}

A total of 949 subjects participated in our experiment. Taken globally, results contain a lot of noise, since only 449 subjects passed the comprehension questions. Here we restrict our analysis to subjects who passed all comprehension questions and we refer the reader to the Supplementary Online Material for the analysis of those subjects who failed the attention check. We include in our analysis also subjects who did not obey the time constraint in order to avoid selection problems that impair causal inference \cite{tinghog2013intuition}.

First we ascertain that our time manipulation effectively worked. Analyzing participants' decision times, we find that those in the time delay condition ($N=246$) took, on average, 45.64 seconds to make their decision, while those under time pressure took, on average, only 20.04 seconds. Thus, although many subjects under time pressure did not obey the time constraint, the time manipulation still had a substantial effect. 

Participants under time pressure transferred, on average, 27.93\% of their endowment, and those under time delay transferred, on average, 28.57\% of their endowment. Linear regression using time manipulation as a dummy variable confirms that the difference is not statistically significant (coeff $=0.00646201$, $p=0.8572$), even after controlling for age, sex, and level of education (coeff $=0.0108648$ $p=0.7265$). We note that restricting the analysis to subjects who obeyed the time constraint leads to qualitatively equivalent results (26.11\% under time pressure vs 29.02\% under time delay, coeff $=0.310051$, $p=0.4868$). Thus, promoting intuition versus reflection does not have any effect on cooperative decision-making. \emph{En passant}, we note that these percentages are far below those observed among US residents in a very similar experiment \cite{capraro2014heuristics}. More precisely, in this latter paper, US residents started out with a $\$0.10$ endowment and were asked to decide how much, if any, to give to the other person. As in the current study, the amount transferred would be multiplied by 2 and earned by the other player. Strictly speaking, these two experiments are not comparable for three reasons. First, in \cite{capraro2014heuristics} there was no time manipulation; second, the initial endowments were different; third, stakes used in the experiment in the US did not correspond to the same stakes in Indian currency. According to previous research, these differences are minor. Indeed, recent studies have argued that stakes do not matter as long as they are not too high \cite{horton2011online,amir2012economic} and that neutrally framed PDs give rise to a percentage of cooperation sitting between that obtained in the time pressure condition and that obtained in the time delay condition \cite{rand2012spontaneous}. Thus comparing the percentage of cooperation in the current study (28\%) with that reported in \cite{capraro2014heuristics} (52\%) supports our assumption that the average Indian sample is particularly non-cooperative or, at least, less cooperative than the average US sample. 

Next we investigate our main research questions. Figure 1 summarizes our results, providing visual evidence that (i) promoting intuition versus reflection has no significant effect on cooperation among inexperienced subjects; that (ii) experienced subjects cooperate more than inexperienced subjects, but only when acting under time pressure.

More specifically, we find that inexperienced subjects under time pressure ($N=47$) transferred, on average, $19.78\%$ of their endowment while those under time delay ($N=73$) transferred, on average, $26.98\%$ of their endowment. The difference is not significant (coeff $=0.0719907$, $p=0.2306$), even after controlling for all socio-demographic variables (coeff $=0.0711474$, $p=0.2337$). Thus, promoting intuition versus reflection has no effect on cooperation among inexperienced subjects living in a non-cooperative setting. This finding is robust to controlling for people who did not obey the time constraint (coeff $=0.156951$, $p=0.5560$) and to controlling for people who obeyed the time contraint (coeff $=0.0371963$, $p=0.8082$).

To explore the interaction between experience and cooperative behavior, as in previous studies \cite{rand2012spontaneous,rand2014socialA,rand2014reflection}, we separate subjects into experienced and inexperienced. This procedure comes from the observation that, although the level of experience is a categorical variable, the association between participant's objective level of experience and their answer to our question is objective only in case of inexperienced subjects. Linear regression predicting cooperation using experience as a dummy variable confirms that experience with economic games on cooperation favors the emergence of cooperative choices, but only among people in the time pressure condition (time pressure: coeff $=1.05974$, $p=0.04464$; time delay: coeff $=0.217945$, $p=0.63930$). These results are robust to including control on the socio-demographic variables (time pressure: coeff $=1.05054$, $p=0.04641$; time delay: coeff $=0.215053$, $p=0.64603$). Our main results are also robust to using non-parametric tests, such as Wilcoxon rank-sum: the rate of cooperation of inexperienced subjects is not statistically distinguishable from the rate of cooperation of inexperienced subjects acting under time delay both when they act under time pressure ($p=0.3271$) and under time delay ($p=0.5619$); and the rate of cooperation of inexperienced subjects is significantly smaller than that of experienced subjects, but only among those acting under time pressure (time pressure: $p=0.0251$; time delay: $p=0.4715$). For completeness, we also report the results of linear regression predicting cooperation using level of experience as independent variable. We find that level of experience has a marginally significant positive effect on cooperation among subjects acting under time pressure (coeff $=0.327449$, $p=0.06343$) and has no effect on cooperation among subjects acting under time delay (coeff = $0.210234$, $p=0.20062$). Also these results are robust to including control on all the socio-demographic variables (time pressure: coeff $=0.305278$, $p=0.08438$; time delay: coeff $=0.19119$, $p=0.25112$).

The increase of cooperation from inexperienced subjects to experienced subjects seems to be driven by participants under time pressure who did \emph{not} obey the time constraint. Specifically, linear regression predicting cooperation using experience as a dummy variable yields non-significant results in case of participants who obeyed the time pressure condition (without control: coeff $=0.137931$, $p=0.8864$; with control: coeff $=0.0760645$, $p=0.9390$) and significant results in case of participants who did not obey it (without control: coeff $=0.151793$, $p=0.0166$; with control: coeff $=0.150522$, $p=0.0180$). This is not surprising and it is most probably due to noise generated by a combination of two factors: the set of people who obeyed the time pressure contraint is very small (for instance, only 14 inexperienced people obeyed the time constraint) and it is more likely to contain people who did not understand the decision problem but passed the comprehension questions by chance (which we estimated to be 5\% of the total. See Supplementary Online Material). 

\begin{figure} 
   \centering
   \includegraphics[scale=0.50]{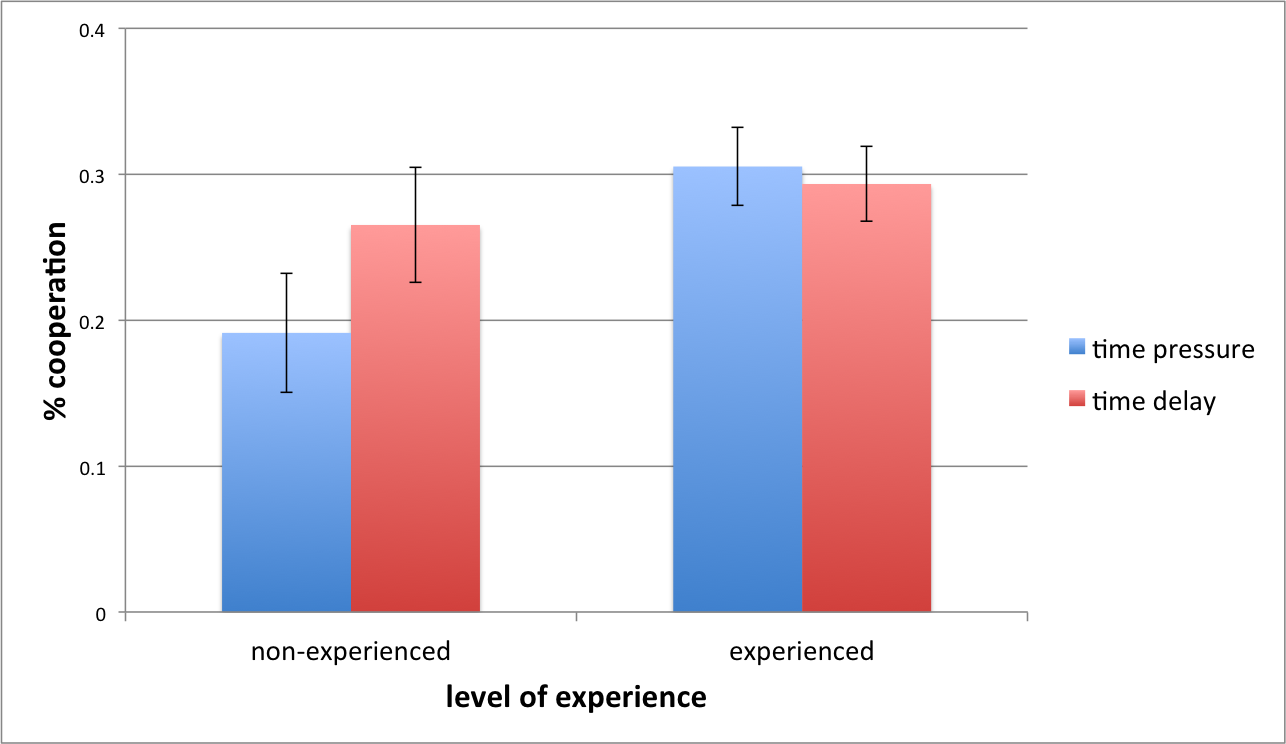} 
   \caption{\emph{Average amount transferred per condition (time pressure vs time delay) divided by level of experience in economic games on cooperation (na\"ive vs non-na\"ive). Error bars represent the standard error of the mean. Inexperienced subjects under time delay transferred slightly more than inexperienced subjects under time pressure, but the difference is not statistically significative ($p=0.2306$). Previous experience with economic games on cooperation has a positive effect on cooperation, but only among participants in the time pressure condition ($p=0.04464$).}}

   \label{fig:aggregated}
\end{figure}

\section*{Discussion}

We have shown that (i) promoting intuition via time pressure versus promoting deliberation via time delay has no effect on cooperative behavior among subjects residents in India with no previous experience with economic games on cooperation, and that (ii) experience has a positive effect on cooperation, but this effect is significant only among subjects acting under time pressure.  

Our results have several major implications, the first of which is providing further support for the Social Heuristics Hypothesis (SHH)  \cite{rand2012spontaneous,rand2014socialA}. Introduced in order to organize the growing body of literature providing direct \cite{rand2012spontaneous,rand2014socialA,cone2014time,rand2014reflection,rand2014socialB,schulz2012affect,cornelissen2011social,roch2000cognitive,lotz2014spontaneous,ruff2013changing} and indirect \cite{engel2014clean,kieslich2014cognitive,righetti2013low,rand2014risking} evidence that, on average, intuitive responses are more cooperative than reflective responses, the SHH contends that people internalize strategies that are successful in their everyday social interactions and then apply them to social interactions that resemble situations they have encountered in the past. Thus, when they encounter a new or atypical situation, people tend to rely on these heuristics and use them as intuitive responses. Deliberation can override these heuristics and adjust the behavior towards one that is more tailored to the current interaction. 

As such, the SHH makes a prediction that has not been tested so far: inexperienced subjects living in a non-cooperative setting should act non-cooperatively both under time pressure, because they use their non-cooperative default strategy (learned in the setting where they live), and under time delay, because defection is optimal in one-shot interactions.  Our results support this prediction.

Besides this prediction, the SHH is also consistent with an interaction between level of previous experience with economic games on cooperation, time pressure, and cooperation in one-shot interactions: experienced people, despite their living in a non-cooperative setting, \emph{might} have internalized a cooperative strategy, to be used only in AMT. The SHH does not predict that a substantial proportion of  experienced people have \emph{in fact} developed this context-dependent intuition for cooperation, but it is certainly consistent with a positive effect of experience on cooperation driven by intuitive responses. Our results provide evidence for this phenomenon.

As mentioned in the Introduction, Kohlberg's rationalist approach makes the explicit prediction that promoting intuition should always undermine cooperation. Thus our results support the SHH versus Kohlberg's rationalistic approach. Of course, this does \emph{not} imply that the rationalist approach should be completely rejected: it is indeed supported by many experimental studies involving pro-social behaviors other than cooperation. If anything, our results point out that different pro-social behaviors may emerge from different cognitive processes. Classifying pro-social behaviors in terms of the processes involved is an important direction for future research towards which, to the best of our knowledge, only one recent study has attempted a first step \cite{corgnet2015cognitive}.

Supporting the SHH, our results suggest that economic models of human cooperation should start taking dual processes and individual history into account. Indeed, virtually all major models of human cooperation are static and decontextualized and only a handful of papers have recently attempted a first step in the direction of taking dual processes into account \cite{fudenberg2006dual,fudenberg2011risk,fudenberg2012timing,dreber2014altruism}. We believe that extending these approaches to incorporate also individual history could be a promising direction for future research.

Our findings go beyond the mere support of the SHH. Our cross cultural analysis, although it is formally not correct, shows that residents in India are, on average, less cooperative than US residents. The difference is so large (28\% vs 52\%) that it is hard to explain it by appealing to minor differences in the experimental designs and so it deserves to be commented.

One possibility, supported by the experimental evidence that good institutions are crucial in promoting cooperation \cite{bardhan2000irrigation,dayton2000determinants,fujiie2005conditions,henrich2005economic,henrich2006costly,herrmann2008antisocial,buchan2009globalization,gachter2009reciprocity,gachter2010culture,buchan2011global,andersson2011inequalities,bigoni2014amoral} and the evidence that India struggles on a daily basis to fight corruption in politics at both the national and local levels  \cite{das2001public,guha2007india,quah2008curbing,miklian2013corruption}, is that residents in India may have internalized non cooperative behavior in their everyday life (because cooperation is not promoted by their institutions) and they tend to apply it also to the new situation of a lab experiment. One far-reaching consequence of this interpretation is that the role of local institutions may go far beyond regularizing behavior. If institutions do not support cooperative behavior, selfishness may even get internalized and applied to atypical situations where people rely on heuristics. While this interpretation is supported by a recent study \cite{peysakhovich2015habits} showing that norms of cooperation learned in one experiment spill over to subsequent experiments where there are no norms, we recommend caution on our interpretation, since our results do \emph{not} show directly that inexperienced residents in India are less cooperative than US residents \emph{because} they are embedded into a society whose institutions do not promote cooperative behavior. However, we believe that this is a fundamental point that deserves to be rigorously addressed in further research.

Interestingly, we have shown that experienced residents in India are significantly more cooperative than inexperienced ones.  This correlation appears to be even more surprising if seen in light of recent studies reporting that experience has a \emph{negative} effect on cooperation among residents in the US \cite{capraro2014heuristics,rand2014socialB}. Although the sign of these effects are different, they share the property that they are driven by intuitive responses. Thus they are in line with the SHH, which assumes that experience operates mainly through the channel of intuition, but it does not make any prediction about the sign of the effect of experience, which may ultimately depend on a number of factors. While it is relatively easy to explain a negative effect of experience with economic games on cooperation, by appealing to learning of the payoff maximizing strategy, explaining a positive effect is harder. One possibility is that experienced subjects have learned cooperation in iterated games, where it might be strategically advantageous, and tend to apply it also in one-shot games. Another possibility is that Turkers are developing a feeling of community that may favor the emergence of pro-social preferences. Understanding what mechanisms can promote the emergence of cooperation from a non-cooperative setting is certainly a fundamental topic for further research.

\pagebreak
\begin{center}
\huge{Supplementary Information}
\end{center}
\wl
This Supplementary Information is divided in three sections: in the first section we report more details about the statistical analysis; in the second section we report full instructions of our experiment. In the third section we report the results of our pilot (containing a methodological error in the measure of participants' level of experience). With some differences in the details, these results are in line with those reported in the Main Text and thus provide further evidence in support of our main result that experienced subjects cooperate more than inexperienced subjects and that experience operates primarily through the channel of intuition.

\section*{More details about the statistical analysis}

\subsection*{Socio-demographics}

We start by reporting the socio-demographics of participants who passed the comprehension questions. Data are summarized in the table. We remind that participants' level of experience was measured using a 5-point Likert-scale from 1=Never to 5=Several times. The socio-demographic statistics show that the majority of subjects in the time pressure condition did not obey the time constraint. This is likely due to the fact that reading the instructions of the decision problem takes about six seconds and so participants had only 4 seconds to understand the problem and make their decision. However, the mean decision time of participants in the time pressure condition was much smaller than the mean decision time of participants in the time delay condition, providing evidence that the time manipulation still had a substantial effect. 

\begin{center}
\begin{tabular}{| l | c | c |}
  \hline
  & pressure & delay\\
  \hline                       
  N & 203 & 246 \\
  sex (M=0, F=1) & 0.3054 & 0.3198 \\
  age & 31.77 & 32.52 \\
  experience & 2.57 & 2.48 \\
  na\"ivety (na\"ive=0,non-na\"ive=1) & 0.6685 & 0.7005\\
  failed time constraint & 64\% & 2\%\\
  decision time (in seconds) & 20.04 & 45.64\\
  \hline  
\end{tabular}\captionof{table}{Socio-demographics of the subjects who participated in our experiment.}
 \end{center}

\subsection*{Regression tables}

In the main text, we have reported  several regression results. Here we summarize the main ones in a table (See Table 2).

\begin{center}
\begin{tabular}{| l | c | c | c | c | c | c |}
  \hline
  & pressure & pressure & pressure & delay & delay & delay\\
  \hline                       
  experience & 1.06** & 1.05** & 0.98* & 0.22 & 0.21 & 0.15\\
  & (0.52) & (0.52) & (0.52) & (0.46) & (0.47) & (0.46)\\
 \hline
  sex & & -0.65 & 0.60 & & -0.07 & -0.06\\
   & & (0.18) & (0.48) & & (0.46) & (0.46)\\
\hline
  age & & -0.00& -0.01 & & 0.01 & 0.01\\
 & & (0.02) & (0.02) & & (0.02) & (0.02)\\
\hline
  education & &  0.39 & 0.40 &  & 0.44 & 0.42\\
 & & (0.32) & (0.32)  & & (0.30)& (0.29)\\
\hline
  time & & & -0.69& &   & 1.80\\
 & & & (0.64) &  & & (1.52)\\
\hline
  constant & 2.96*** & 1.85& 1.99 & 3.73*** & 1.36 & −0.38\\
  & (0.46) & (1.91) & (1.91) & (0.39) & (1.89)& (2.39)\\
\hline
  No. cases & 203 & 203 &203& 246 & 246&246\\
  \hline  
\end{tabular}\captionof{table}{Summary of the statistical analysis. For each of the two conditions (pressure and delay) we report the results of linear regression using experience as a dummy variable, with and without control on all the demographic variables and on dummy variable ``time'', which is 1 is the subject obeyed the time constrain and 0 otherwise. We report coefficient, standard error, and significance levels, using the notation: *: $p<0.1$, **: $p<0.05$, and ***: $p<0.01$.}
 \end{center}

\subsection*{Estimation of the percentage of subjects who failed the attention check}

Finally we analyze subjects who failed the attention test. Indeed, the large number of participants who failed the comprehension questions (about half of the total number) is worrisome that a substantial fraction of those who passed the comprehension questions may have passed them just by chance. Although this rate of passing is not much lower than that in similar studies conducted in the US (in a very similar experiment, conducted with American subjects and published in \cite{capraro2014heuristics}, 32\% of subjects did not pass the comprehension questions) it could potentially generate noise in our data. Indeed, subjects who failed the attention test played essentially at random (time pressure: average transfer = 47.20\%; time delay: average transfer = 46.35\%).

To exclude this possibility, we analyse failers' responses to the comprehension questions. We start with the first comprehension question, which asked ``What is the choice by YOU that maximizes YOUR outcome?''. Participants could choose any even amount of money from 0c to 20c, for a total of 11 possible choices. Figure 1 reports the distribution of responses of people who failed at least one comprehension questions. The distribution is clearly tri-modal, with 60\% of responses equally distributed among the two extreme choices (full cooperation and full defection) and the mid-point (transfer half). The remaining 40\% is equally distributed among all other choices. Consequently, assuming that confused participants respond to the first comprehension question according to this distribution, the probability that a confused participant pass the first comprehension question by chance is equal to 1/5. 

\begin{figure} 
   \centering
   \includegraphics[scale=0.90]{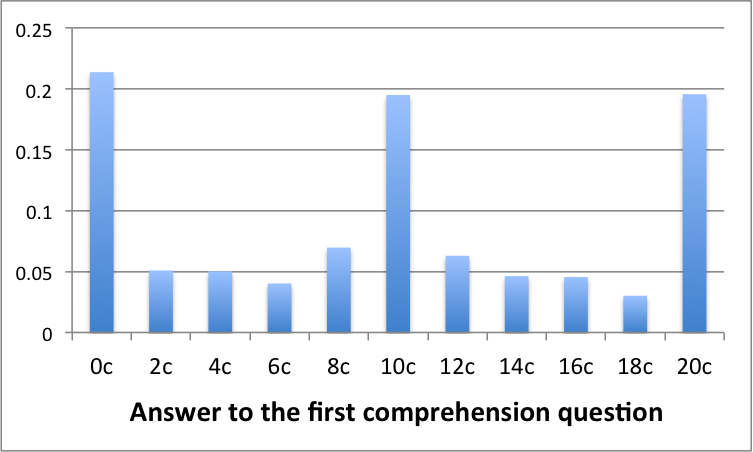} 
   \caption{\emph{Distribution of responses to the first comprehension question from participants who failed at least one comprehension question. The distribution is clearly tri-modal, with about 60\% of responses equally distributed among the two extreme choices and the mid-point. The remaining 40\% is equally distributed among all other choices.}}
\label{fig:aggregated}
\end{figure}

Next we analyze the responses to the second comprehension question, which asked ``What is the choice by YOU that maximizes the OTHER PARTICIPANT's outcome?''. Figure 2 reports the distribution of answers. Also in this case we find a tri-modal distribution, although this time the correct answer appeared with higher frequency (about 50\%). Assuming that a confused participant answered this question according to this distribution, we conclude that a confused participant had probability about 1/10 to pass the first two comprehension questions by chance.

\begin{figure} 
   \centering
   \includegraphics[scale=0.90]{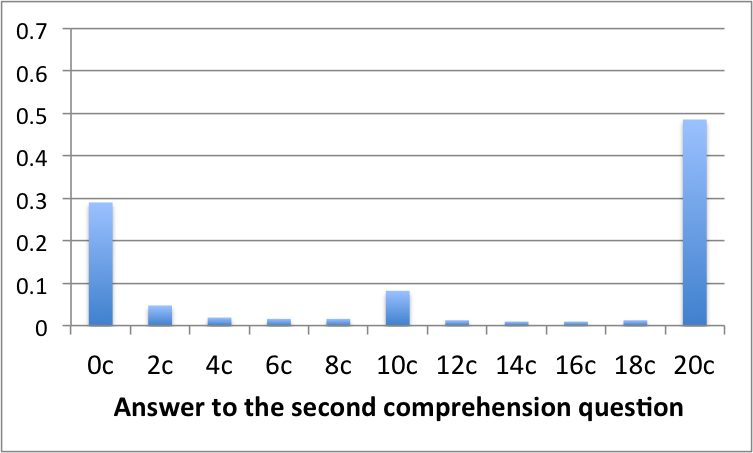} 
   \caption{\emph{Distribution of responses to the second comprehension question from participants who failed at least one comprehension question. About half of these participants answered correctly.}}
\label{fig:aggregated}
\end{figure}

Then we analyze the responses to the third comprehension question, which asked ``What is the choice by the OTHER PARTICIPANT that maximizes YOUR  outcome?'' Figure 3 reports the distribution of answers. This time the distribution is essentially bi-modal, with about 3/5 of people answering correctly. Assuming that confused participants answered according to this distribution, we conclude that a confused participant had probability about 3/50 to pass the first three comprehension questions by chance.

\begin{figure} 
   \centering
   \includegraphics[scale=0.90]{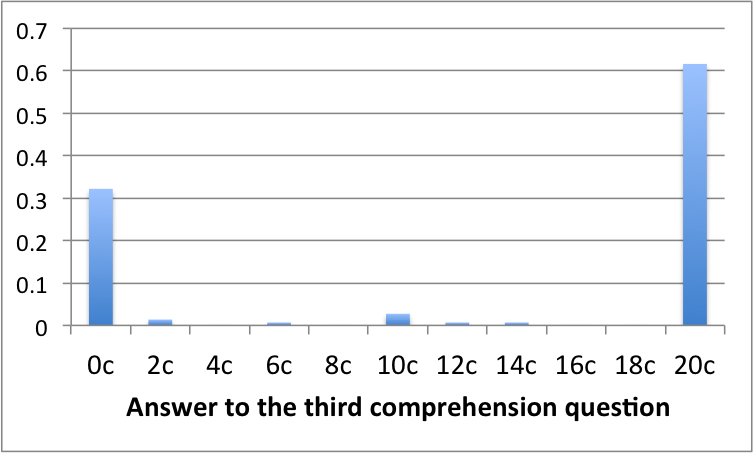} 
   \caption{\emph{Distribution of responses to the second comprehension question from participants who failed at least one comprehension question. About 3/5 of these participants answered correctly.}}
\label{fig:aggregated}
\end{figure}

Now, it is impossible to have a clue of what the proportion of confused participants who passed the fourth comprehension question by chance is. The analysis above suggests that 3/50 is an upper bound of the probability that a confusing subject passed all comprehension questions by chance. Interestingly, 88\% of subjects who answered 20c in the third comprehension and failed the fourth comprehension question, answered 20c also to the fourth comprehension question, which asked ``What is the choice by the OTHER PARTICIPANT that maximizes the OTHER PARTICIPANT's outcome?''. This suggests that the actual probability of passing all comprehension questions by chance is much lower that 3/50. In any case, although our results do not allow to make a precise estimation of noise, the proportion of people who passed the comprehension questions by chance is likely below 5\% of the total, suggesting that noise is a minor problem in our data.

\section*{Experimental instructions}

Participants were randomly assigned to either the time pressure condition or the time delay condition. In both conditions, after entering their worker ID, participants were informed that they would be asked to make a choice in a decision problem to be presented later and that comprehension questions would be asked. Participants were also informed that the survey (which was made using the software Qualtrics) contained a skip logic which would automatically exclude all participants failing any of the comprehension questions. Specifically, this screen was as follows:

\emph{Welcome to this HIT.} 
 
\emph{This HIT will take about five minutes. For the participation to this HIT, you will earn 0.50 US dollars, that is, about 31 INR. You can also earn additional money depending on the decisions that you and the other participants will make.}

\emph{You will be asked to make one decision. There is no incorrect answer. However:}

\emph{IMPORTANT: after making the decision, to make sure you understood the decision problem, we will ask some simple questions, each of which has only one correct answer.  If you fail to correctly answer any of those questions, the survey will automatically end and you will not receive any redemption code and consequently you will not get any payment.}

\emph{With this in mind, do you wish to continue?}

At this stage, they could either leave the study or continue. Those who decided to continue were redirected to an introductory screen where we gave them all the necessary information about the decision problem, but without telling exactly which one it is. This is important in order to have the time pressure and time delay conditions work properly in the next screen. This introductory screen for the participants in the time pressure condition was the following:

\emph{You have been paired with another participant. You can earn additional money depending on the decision you will make in the next screen. You will be asked to make a choice that can affect your and the other participant's outcome. The decision problem is symmetric: also the other participant is facing the same decision problem. After the survey is completed, you will be paid according to your and the other participant's choices.}

\emph{You will have only 10 seconds to make the choice.}

\emph{This is the only interaction you have with the other participant. He or she will not have the opportunity to influence your gain in later parts of the HIT. If you are ready, go to the next page.}

The introductory screen for the participants in the time delay condition was identical, a part from the fact that the sentence `You will have only 10 seconds to make the choice' was replaced by the sentence `You will be asked to think for at least 30 seconds before making your choice. Use this time to think carefully about the decision problem'.

The decision screen was the same in both conditions:

\emph{You and the other participant are both given $\$0.20$ US dollars. You and the other participant can transfer, independently, money to the each other. Every cent you transfer, will be multiplied by $2$ and earned by the other participant. Every cent you do not transfer, will be earned by you.}

\emph{How much do you want to transfer?}

By using appropriate buttons, participants could transfer any even amount of money from $\$0$ to $\$0.20$. 

En passant, we observe that reading the decision screen takes about six seconds and thus participants under time pressure had only about four seconds to make their choice.

To assure that time pressure and time delay work properly, it is necessary that comprehension questions are asked after the decision has been made. Thus, right after the decision screen, participants faced the following four comprehension questions. 

\emph{What is the choice by YOU that maximizes YOUR outcome?}

\emph{What is the choice by YOU that maximizes the OTHER PARTICIPANT's outcome?}

\emph{What is the choice by the OTHER PARTICIPANT that maximizes YOUR  outcome?}

\emph{What is the choice by the OTHER PARTICIPANT that maximizes the OTHER PARTICIPANT's outcome?}

By using appropriate buttons, participants could select any even amount of money from $\$0$ to $\$0.20$. Participants who failed any of the comprehension questions were automatically excluded from the survey. Those who answered all questions correctly entered the demographic questionnaire, where we asked for their age, sex, reason for their choice, and, most importantly, level of experience in these games. Specifically, we asked the following question:

\emph{To what extent have you previously participated in other studies like to this one (e.g., exchanging money with strangers)?}

Answers were collected using a 5 point Likert-scale from ``1=Never'' to ``5=Several times''.

\section*{Experimental results of the pilot}

Our pilot experiment was identical to our main experiment, except for the fact that, as a measure of experience, we asked participants to self-report the extent to which they had participated in ``exactly'' the same task before. Participants could choose between: never, once or twice, and several times. The use of the word ``exactly'' is problematic, since it might lead to confusion: what does the answer ``exactly the same task'' imply? Does it imply that participants have participated in a task with exactly the same instructions (including time constrains) or does it imply that the participant have participated in a task containing the same economic game? Figure 4 reports the results of the pilot. We observe that, indeed, the details are different: level of experience seem to have a inverted-U effect on cooperation, which, since it has not been replicated in the main experiment, is probably due to confusion regarding the interpretation of the word ``exactly'' . However, as in our main experiment, we find that experienced subjects are significantly more cooperative than little or no experienced subjects and that this behavioral change is mainly driven by intuitive responses (see Figure 5). 

\begin{figure} 
   \centering
   \includegraphics[scale=0.60]{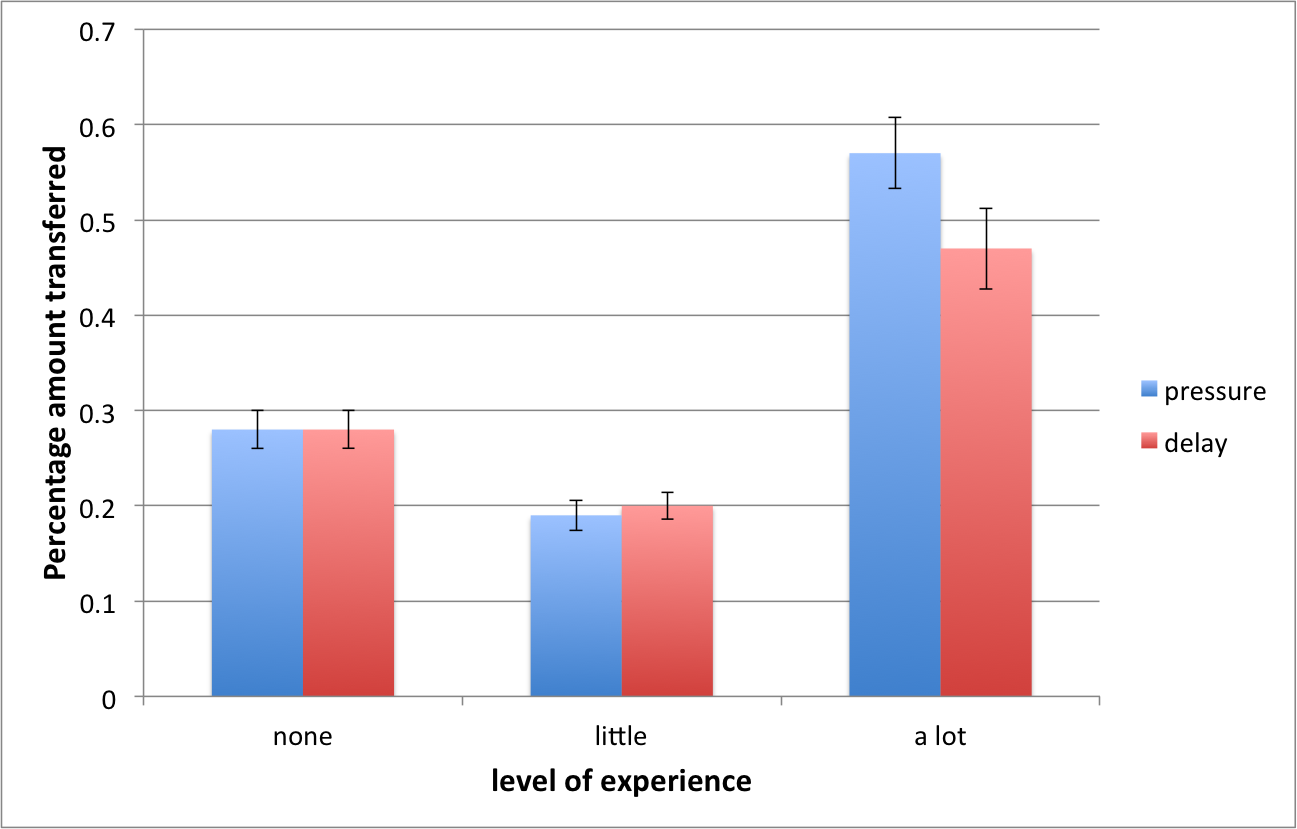} 
   \caption{\emph{The figure reports the percentage of the endowment transferred as a function of the level of experience and condition (time pressure versus time delay). Error bars denote the standard error of the mean. Promoting intuition versus reflection does not have any effect among little or no experienced subjects. On the contrary, it seems to have a positive effect on cooperation among experienced subjects. Linear regression shows that this effect is nearly significant. Level of experience seems to have a U-shaped effect on cooperation. Linear regression confirms that little experienced subjects are significantly less cooperative than both inexperienced and experienced subjects, and that experienced subjects are significantly more cooperative than inexperienced subjects. As we have shown in the Main Text, the particular inversed-U effect is probably due to the confusion generated by the use of the word ``exactly'' in the measure of participants' level of experience. The fact that experience promotes cooperation by changing intuitive responses more than reflective responses is in line with the results reported in the Main Text.}}
\label{fig:aggregated}
\end{figure}

Specifically, linear regression confirms that little experienced subjects are significantly less cooperative than inexperienced subjects both under time pressure (coeff $= -0.0844337$, $p = 0.00088$) and under time delay (coeff $= -0.085229$, $p = 0.00031$); and confirms that experienced subjects are significantly more cooperative than little experienced subjects both under time pressure (coeff $= 0.374666$, $p < .0001$) and under time delay (coeff $= 0.272971$, $p < .00001$). Moreover, experienced subjects were also significantly more cooperative than inexperienced subjects, both under time pressure (coeff $= 0.145116$, p $< .0001$) and under time delay (coeff $= 0.0938709$, p $< .0001$). Thus experience has a significant inverted-U effect on cooperation, where little experienced subjects cooperate the least and experienced subjects the most.
The coefficients of the previous regressions suggest that the motivations behind the initial decrease of cooperation, which affects subjects under time pressure and those under time delay to exactly the same extent, are different from the motivations behind the subsequent flourishing of cooperation, which seems to affect subjects under time pressure to a larger extent than those under time delay. To confirm this, we use linear regression to predict decision among experienced subjects using time pressure as a dummy variable. We find that experienced subjects under time pressure are nearly significantly more cooperative than experienced subjects under time delay (coeff $= 0.0960161$, $p = 0.09165$).

\end{document}